\input{aipcheck}

\documentclass[
    ,final            
  ]
  {aipproc}

\layoutstyle{8x11single}

\begin{document}

\title{The Sanford Underground Research Facility at Homestake}


\author{J.\ Heise}{
  address={Sanford Underground Research Facility, 630 East Summit Street, 
Lead, SD 57754}
}

\begin{abstract}
The former Homestake gold mine in Lead, South Dakota is being transformed 
into a dedicated laboratory to pursue underground research in rare-process 
physics, as well as offering research opportunities in other disciplines 
such as biology, geology and engineering.  A key component of the Sanford 
Underground Research Facility (SURF) is the Davis Campus, which is in 
operation at the 4850-foot level (4300 m.w.e) and currently hosts three 
projects: the LUX dark matter experiment, the {\sc Majorana Demonstrator} 
neutrinoless double-beta decay experiment and the CUBED low-background 
counter.  Plans for possible future experiments at SURF are well underway 
and include long baseline neutrino oscillation experiments, future dark 
matter experiments as well as nuclear astrophysics accelerators.  
Facility upgrades to accommodate some of these future projects have 
already started.  SURF is a dedicated facility with significant expansion 
capability.
\end{abstract}

\maketitle


\section{Introduction}

Many disciplines benefit from access to an underground facility dedicated 
to scientific research, including physics, biology, geology, engineering, 
and a well-established science program is currently underway at the 
Sanford Underground Research Facility (SURF).  The unique underground 
environment at SURF allows researchers to explore a host of important 
questions regarding the origin of life and its diversity, mechanisms 
associated with earthquakes and a number of engineering topics.  A deep 
underground laboratory is also where some of the most fundamental topics 
in physics can also be investigated, including the nature of dark matter, 
the properties of neutrinos and the synthesis of atomic elements within 
stars.

SURF is being developed in the former Homestake gold mine, in Lead, South 
Dakota~\cite{SURF-Lesko, SURF-Heise}. Barrick Gold Corporation donated the 
site to the State of South Dakota in 2006, following over 125 years of 
mining~\cite{Mitchell-2009}, during which time over 600 km of tunnels and 
shafts were created in the facility, extending from the surface to over 
8,000 feet below ground.  The Laboratory property comprises 186 acres on 
the surface and 7,700 acres underground, and the Surface Campus includes 
approximately 253,000 gross square feet of existing structures. The South 
Dakota Science and Technology Authority (SDSTA) operates and maintains the 
Sanford Laboratory at the Homestake site in Lead, South Dakota with 
management oversight by Lawrence Berkeley National Laboratory (LBNL).

\begin{figure}
  \includegraphics*[scale=0.38]{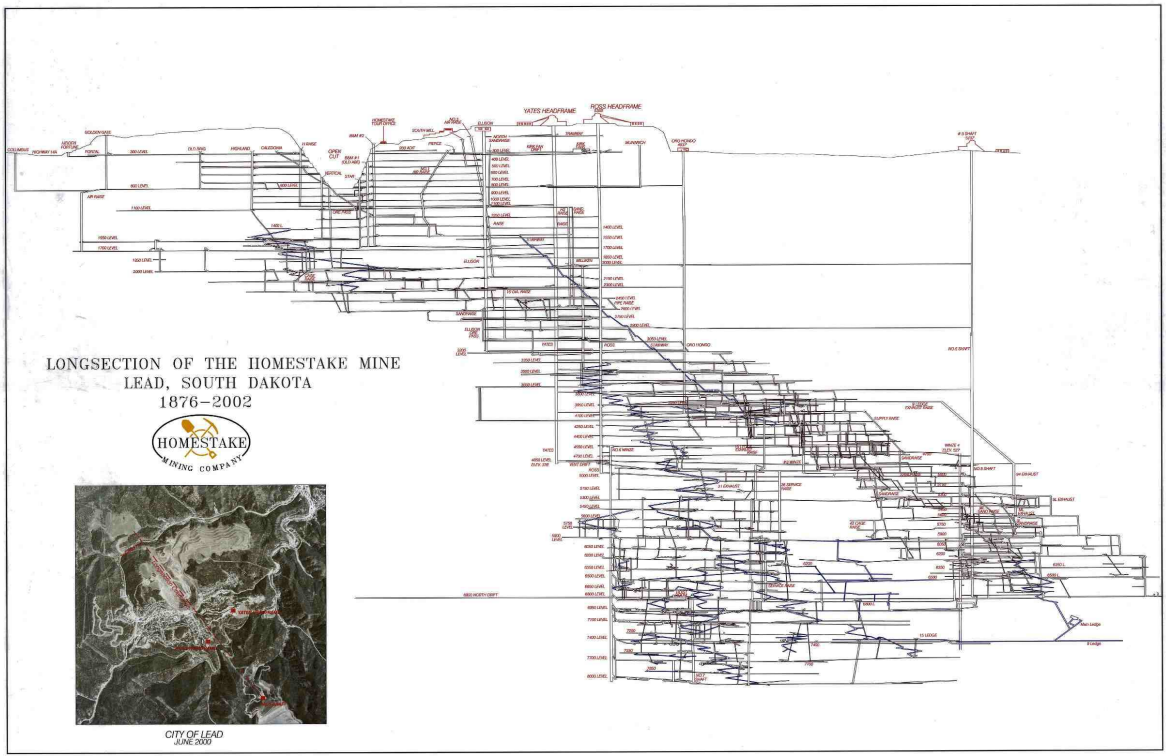}
  \caption{\label{fig:Longsection} The long section of the former 
Homestake Gold Mine. This figure illustrates the 60 underground levels 
extending to greater than 8,000 feet below ground. The location of cross 
section is indicated in the inset along a NW to SE plane. The projection 
extends for 5.2 km along this plane.}
\end{figure}

In 2006, South Dakota philanthropist, T.\ Denny Sanford, gifted \$70M to 
convert the former mine into a research laboratory and develop a science 
education facility. With these funds and a strong commitment from the State 
of South Dakota (appropriations of \$42.2M to date), safe access to the 
underground has been reestablished.

The initial concepts for SURF were developed with the support of the US 
National Science Foundation (NSF) as the primary site for the NSF's Deep 
Underground Science and Engineering Laboratory (DUSEL)~\cite{DUSEL-Lesko}. 
With the National Science Board's decision to halt development of a 
NSF-funded underground laboratory, the US Department of Energy (DOE) now 
supports the majority of the operation of the facility. Support for 
experiments at SURF comes from both the NSF and DOE as well as other 
agencies such as the USGS and NASA.  Elements of the Homestake DUSEL 
Preliminary Design Report~\cite{DUSEL_PDR-Lesko} continue to be useful as 
the feasibility for portions of the original plan are investigated.


\section{Facility Operations Infrastructure}

Maintenance and operation of key elements of facility infrastructure 
enables safe access underground.  Transportation of personnel and 
materials underground is accomplished using the two primary shafts, the 
Ross Shaft and the Yates Shaft.  Pairs of hoists near both the Ross and 
Yates shafts move personnel and rock conveyances through the respective 
shafts.  Pumping stations in the Ross Shaft allow ground water to be 
pumped to surface.  Underground ventilation is provided by the Oro Hondo 
fan as well as the fan at \#5 Shaft, which bring fresh air underground via 
the Ross and Yates Shafts.

A key feature of the Sanford Laboratory is the capacity for redundancy.  
Redundant power, fiber optics cable and ventilation are brought 
underground via the Ross and Yates Shafts.  As well, the two shafts as 
well as numerous ramp systems provide multiple options for emergency 
egress.

Initial rehabilitation of the surface and underground infrastructure 
focused on the Ross shaft that provides access to the majority of 
underground utilities, including the pumping system used to remove water 
from the mine.  Contractor work in the Ross shaft started July 2008 and 
was completed by October 2008, after which SDSTA personnel continued with 
maintenance and renovations, including the removal of unused legacy piping 
(12 km) and power/communication cables (2 km).  Started in November 2008, 
the Yates shaft renovation was completed in May 2012 with the installation 
of a new personnel conveyance and emergency braking system.  The Ross 
Shaft provided primary underground access from the start of re-entry in 
2007 until the Yates shaft was ready in May 2012.  Personnel and materials 
are transported underground mainly using the Yates cage, which has 
dimensions 1.4 m wide $\times$ 2.7 m tall $\times$ 3.8 m long and has a 
maximum load capacity of 10,000 lbs (4,536 kg).  The Yates cage schedule 
accommodates three shifts per day for science personnel, providing 24-hour 
access as needed.

The average ground water inflow into the underground workings is 
approximately 730~gpm.  After pumping ceased in June 
2003~\cite{Murdoch-2012, Zhan_Duex-2010}, the mine filled with water until 
a high-water mark of 1381 meters (4530 feet) below surface was reached in 
August 2008.  Sustained pumping resumed in June 2008, dropping the water 
level below the 4850 Level by May 2009.  Since April 2012 the water level 
is being maintained around the 6000-foot level below surface.  While the 
potential for accommodating deeper access exists, without a funded mandate 
to develop laboratory space below the 4850-foot level, there is benefit in 
terms of cost and safety for maintaining the water level around the 
current level.

A deep-well pump was installed July 2010 and is currently located about 
6424 feet below surface in \#6 Winze, which extends from the 4850 Level to 
the deepest areas of the facility.  Permanent stations employing 
700-horsepower pumps are located on the 1250, 2450, 3650 and 5000 Levels in 
the Ross shaft.  Water received at the surface Waste Water Treatment Plant 
(WWTP) is combined with Homestake-Barrick water and treated to remove iron 
that has leached from the mine workings and trace amounts of ammonia.  The 
discharge capacity from the WWTP is roughly 2000 gpm using biological and 
sand-filter technologies.

Single-mode fiber optics cable is deployed throughout the facility and 
current network hardware provides inter-campus communication at 100--1000 
Mbps.  Redundant connections exist to the outside world, including 
commodity internet (``Internet 1'') at 1~Gbps and research internet 
(``Internet 2'' via the state Research, Education and Economic Development 
(REED) network) at 1~Gbps, which can be expanded to 100~Gbps with 
appropriate hardware upgrades.  Redundant fiber connections connect the 
surface to the 4850L via the Ross and Yates Shafts, and work in underway 
to ensure that the core network equipment is protected by uninterruptible 
power supplies and in most cases generator power.

In order to provide increased capacity to support the construction and 
operation of future experiments, state and private funds have been 
allocated to perform extensive renovations in the Ross Shaft.  
Refurbishment of the Ross Shaft began in August 2012, and over 1100 feet 
of new steel and associated ground support has been installed as of 
December 2013.  A total of 5159 feet of new steel sets will be installed 
as part of the Ross shaft upgrade project, which is expected to be 
completed by mid-2017.  Once the Ross shaft is completed, a planned 
renovation of the Yates Shaft will commence.


\section{Science Facilities}

In addition to the considerable underground extent, several dedicated 
facilities exist at SURF to support science activities.  On the surface, a 
number of spaces are being used for storage (including drill core) and 
limited construction activities.  Three main physics laboratory spaces are 
available: the Surface Laboratory and two campuses on the 4850 Level -- 
the Ross Campus and the Davis Campus.

\subsection{Surface Facilities}

\begin{figure}
\includegraphics*[scale=0.60]{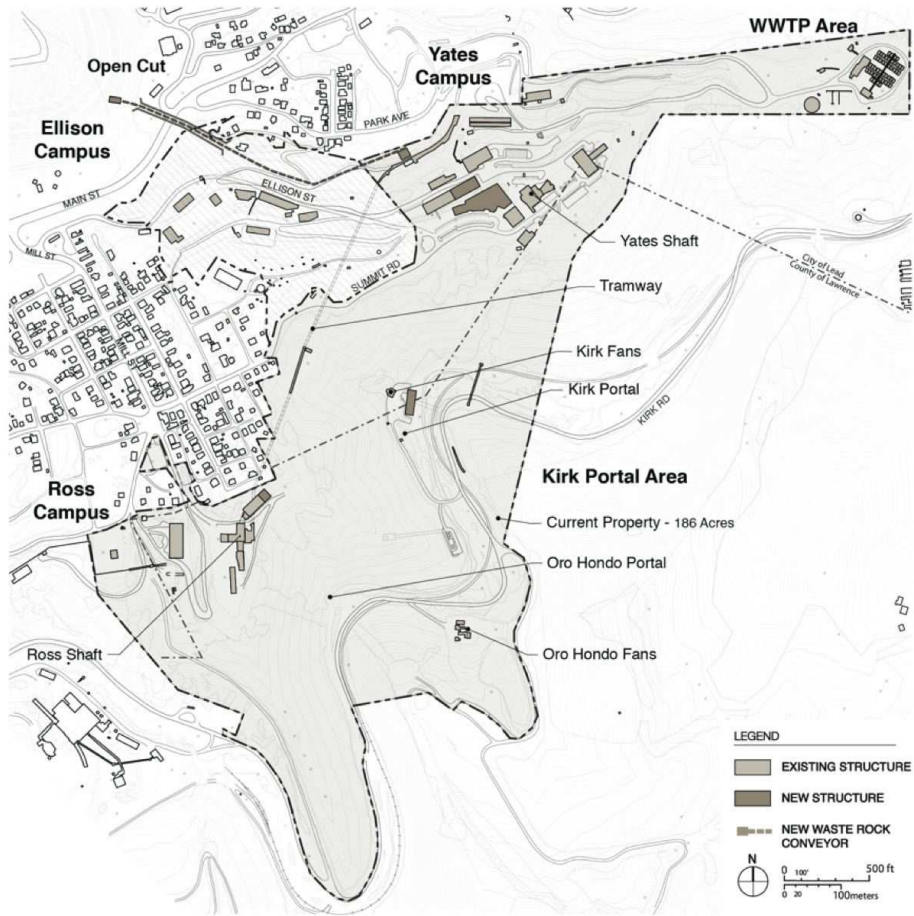}
\caption{\label{fig:SurfaceCampus} A plan view of the surface campus at 
SURF, which comprises 186 acres.}
\end{figure}

Key surface facilities at SURF support science activities at both the 
Yates and Ross surface campuses, including administrative activities, 
education and public outreach functions, the Waste Water Treatment Plant 
(WWTP) and a warehouse for shipping/receiving.  
Figure~\ref{fig:SurfaceCampus} shows the extent of the surface property.

While there are many possibilities of surface buildings that could be 
renovated to directly support science activities, two facilities currently 
function in that capacity: the Core Archive and the Surface Laboratory.

\subsubsection{Core Archive}

Donated by Homestake-Barrick, SURF is the steward of 39,760 boxes of drill 
core, which corresponds to a total length of 91 km.  Homestake core holes 
extend to 10,800 feet below surface.  An additional 5,400 feet of core 
were added to the collection as part of the geotechnical investigations on 
the 4850L for DUSEL.

The SD Geological Survey has assisted with the development of an online 
database that so far includes 58,000+ entries, representing 1,740 drill 
holes.  

\subsubsection{Surface Laboratory}

\begin{figure}
\includegraphics*[scale=0.16]{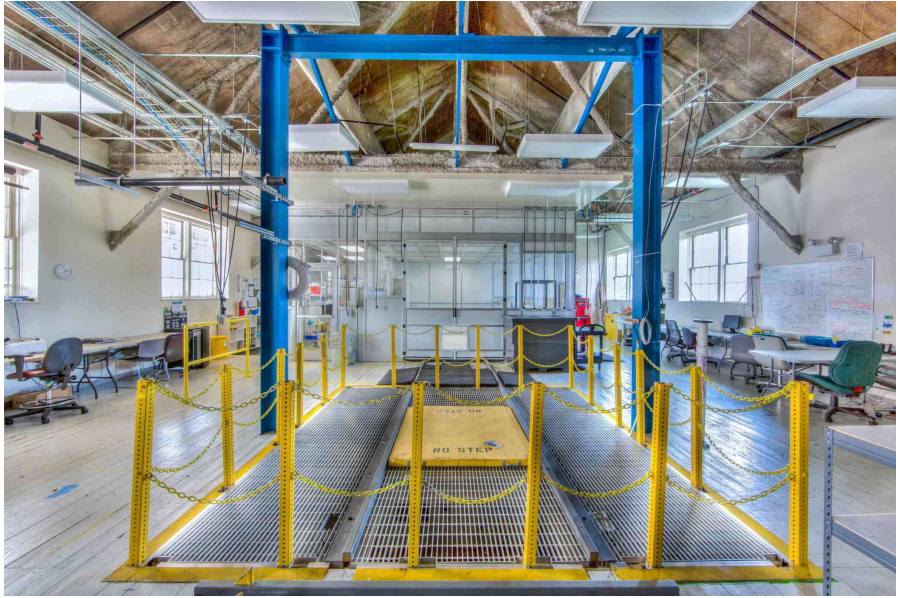}
\caption{\label{fig:SurfaceLab} Surface Laboratory. The recessed area 
with the tank is under the grating in the foreground and the cleanroom is 
located in the background.  A hoist can be installed on the blue I-beam 
structure.} 
\end{figure}

Renovations were undertaken in 2009 in order to transform a former 
warehouse into a laboratory.  Construction was completed in early 2010, 
resulting in approximately 190 m$^{2}$ of lab space in the top-most level 
of a four-storey building as shown in Figure~\ref{fig:SurfaceLab}.  The 
facility includes a cleanroom (5.6 m $\times$ 6.6 m with a 2.7 m ceiling 
height) and corresponding dedicated air handling and filtration system as 
well as a tank that can be used as a water shield (2.8 m diameter $\times$ 
4 m high).  The tank is installed in a recessed shaft in the center of the 
laboratory space.  The laboratory was initially designed to meet the needs 
of the LUX experiment, but is now used to support multiple research 
groups.

\subsection{Underground Facilities}

A number of levels required to support facility operations infrastructure 
can also accommodate research activities. However, the main infrastructure 
for the support of science activities has been developed on the 4850L with 
formal campuses located near both the Ross and Yates shafts.

\subsubsection{4850L Ross Campus}

\begin{figure}
\includegraphics*[scale=0.50]{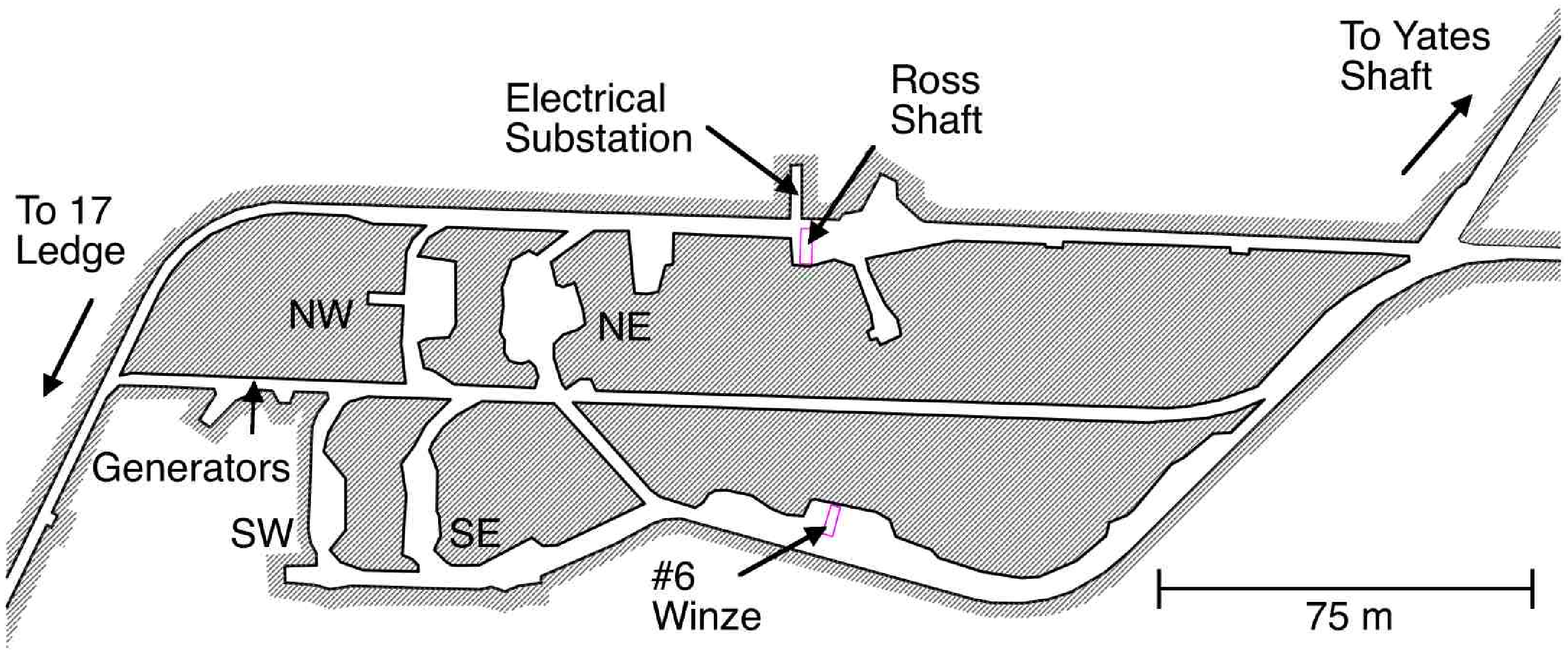}
\caption{\label{fig:RossCampus} 4850L Ross Campus.  Four existing 
excavations are labeled: NW, NE, SE, SW.}
\end{figure}

The 4850L Ross Campus includes the Ross Shaft and \#6 Winze and 
encompasses a set of four existing excavations that were used as 
maintenance shops during mining activities.  These former shops afford an 
economical means to implement experiments or other equipment in a timely 
manner.  The layout is shown in Figure~\ref{fig:RossCampus}, including the 
location of the current electrical substation and generators.


\begin{table}[htbp]
\begin{tabular}{lcc}
\hline \hline
{\bf Ross Campus Location} & {\bf Area}      & {\bf Volume}  \\
                           & {\bf (m$^{2}$)} & {\bf (m$^{3}$)} \\
\hline
NW ({\sc Majorana} Electroforming)  & 184   &  504 \\ 
NE                                  & 297   &  707 \\ 
SE                                  & 137   &  376 \\ 
SW (Refuge Chamber)                 & 121   &  332 \\ 
\hline \hline
\end{tabular}
\caption{\label{tab:RossCampusVol}Footprint areas and volumes for the 
4850L Ross Campus spaces.  The volume for the NE area is derived from 
laser-scan data.}
\end{table} 

\begin{figure}
\includegraphics*[scale=0.50]{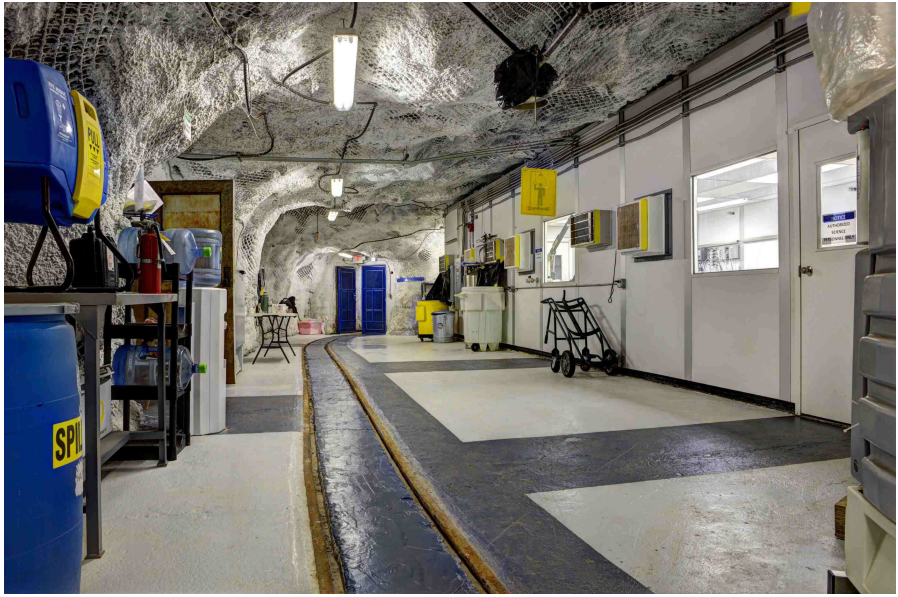}
\caption{\label{fig:RossCampus-MJD} The {\sc Majorana} electroforming 
laboratory at the 4850L Ross Campus.  The cleanroom is located on the 
right-hand side.}
\end{figure}

Two of the shop areas are currently in use.  The SW area has been 
converted into a safety Refuge Chamber that can accommodate 72 people for 
up to 96 hours.  It includes compressed air, CO$_{2}$ scrubbers, air 
conditioners, rations and communications. The NW area is currently being 
used by the {\sc Majorana} collaboration to electroform copper as shown 
in Figure~\ref{fig:RossCampus-MJD}.

The NE shop will house the CASPAR accelerator and a cleanroom is proposed 
for the SE area to host a suite of low-background counters and possibly 
support other research activities, too.

\subsubsection{4850L Davis Campus}

\begin{figure}
\includegraphics*[scale=0.216]{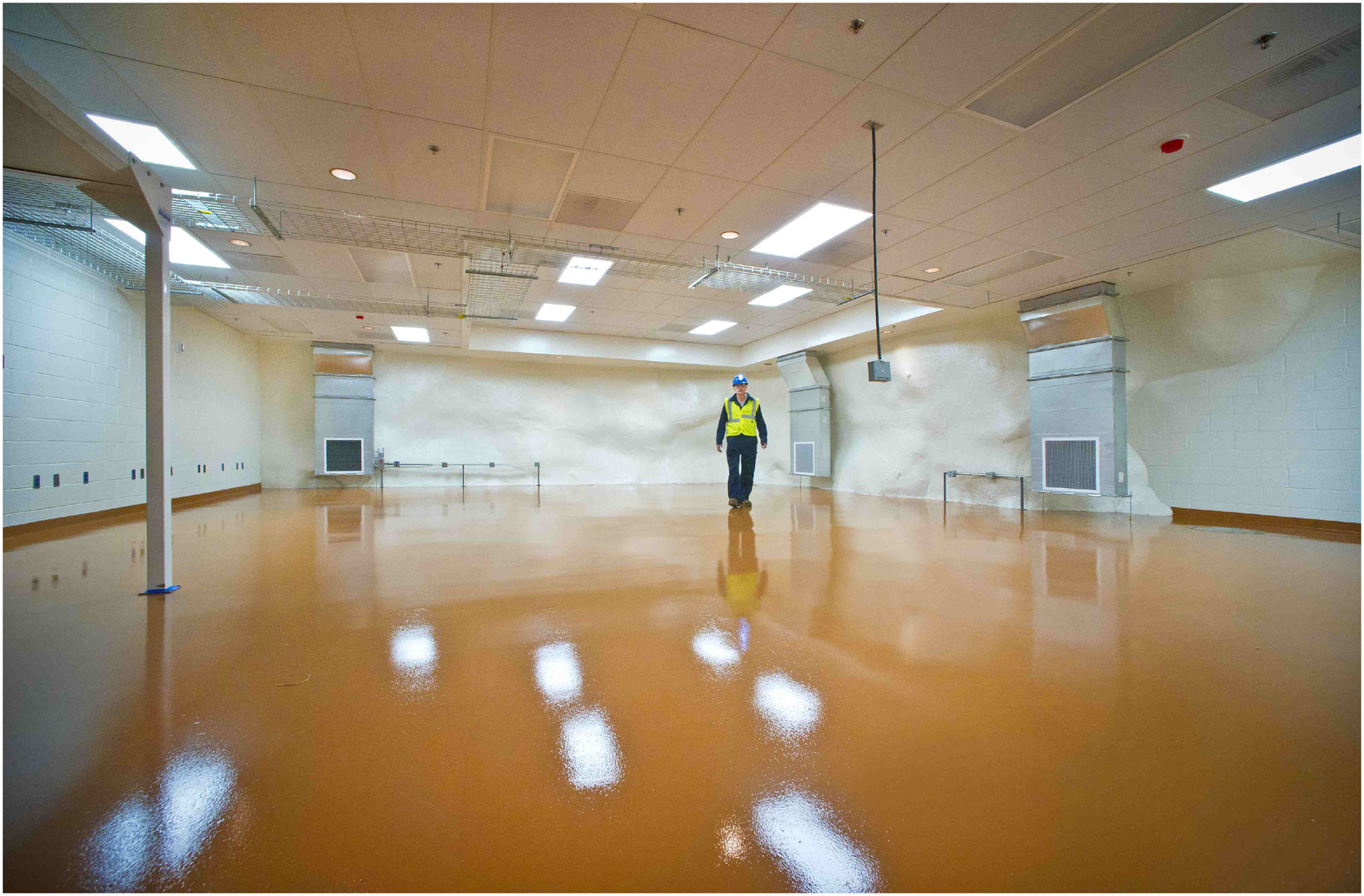} \\ 
\includegraphics*[scale=0.20]{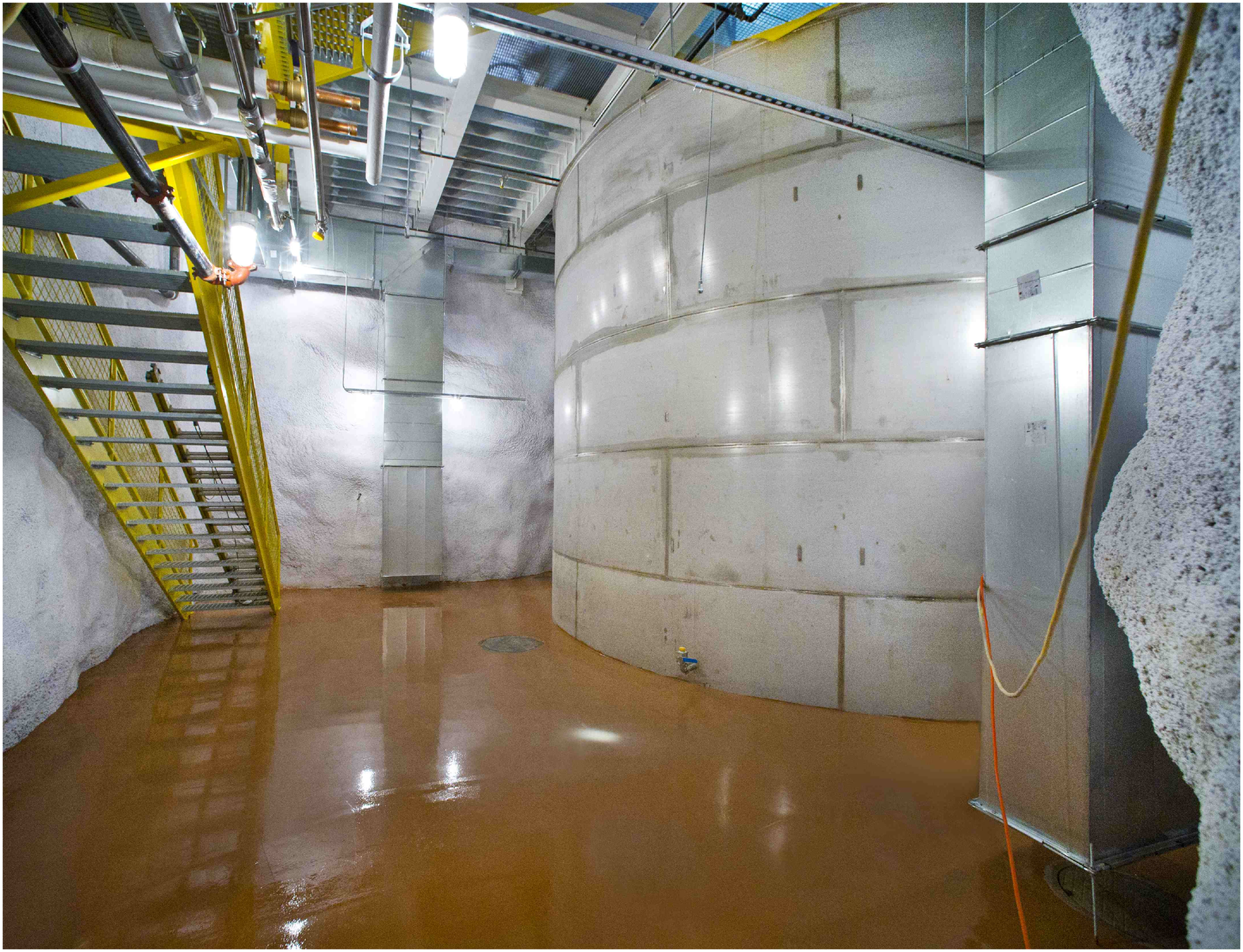}
\caption{\label{fig:DavisCampusPics}Pictures of the Davis Campus in May 
2012 after outfitting had finished: (upper) {\sc Majorana} Detector Room, 
(lower) water shielding tank in Lower Davis for the LUX detector.  A 
considerable amount of equipment has since been installed by both 
collaborations.} 
\end{figure}

A state-of-the-art laboratory complex called the Davis Campus has been 
constructed at the 4850L near the Yates Shaft.  The Davis Campus 
represents a \$15.2M South Dakota commitment using state and private 
funds.  New excavation for the Davis Campus took place during September 
2009 -- January 2011, during which time 16,632 tonnes (18,334 tons) of 
rock was removed.  Rather than being transported to the surface, areas 
were identified on the 5000L (via the 4850L) for rock storage.  Shotcrete 
was applied in both the Davis Cavern (average thickness 12.7~cm) and the 
Transition space (average thickness 8.9~cm).  Laboratory outfitting began 
in June 2011 and was substantially complete in May 2012; pictures are 
shown in Figure~\ref{fig:DavisCampusPics}.  A final occupancy certificate 
was issued in July 2012, which specifies the maximum number of occupants 
to be 62 persons.

\begin{figure}
\includegraphics*[scale=0.53]{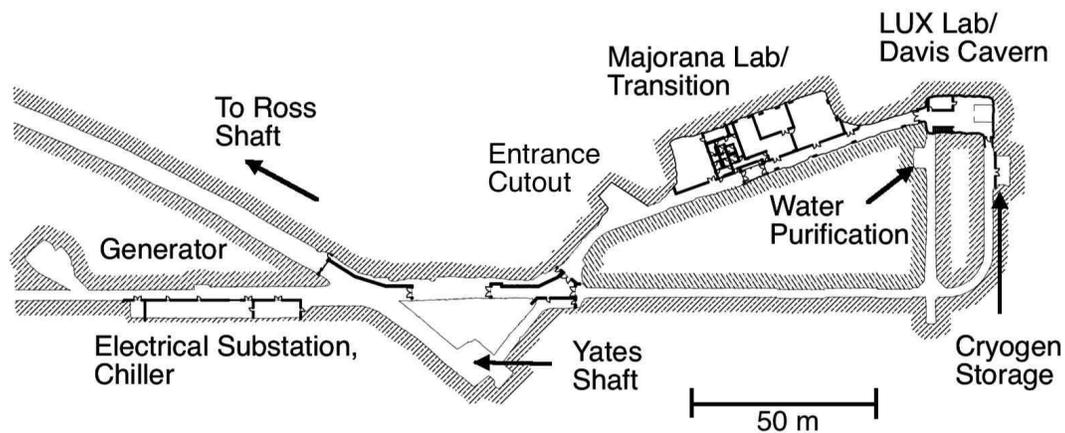}
\caption{\label{fig:DavisCampus} 4850L Davis Campus showing the two main 
laboratory spaces and the proximity to the Yates Shaft.}
\end{figure}

The two main experiments areas are the {\sc Majorana} Lab/Transition space 
(43m L $\times$ 16m W $\times$ 4m H; 3,648 m$^{3}$) and the Davis Cavern 
(17m L $\times$ 10m W $\times$ 12m H; 2,089 m$^{3}$), as shown in 
Figure~\ref{fig:DavisCampus}.  Quoted dimensions are average values based 
on post-shotcrete laser-scan data.  The total Davis Campus footprint 
consists of 927 m$^{2}$ of laboratory space and 2,732 m$^{2}$ total space.


An array of extensometers is installed to monitor ground motion and 
convergence at the Davis Campus -- data collected so far indicate nothing 
unexpected.  Geotechnical interpretations were performed using drill core 
and other rock mechanics studies~\cite{Pariseau-2012}.

Services provided within the Davis Campus include fire sprinklers and 
alarms throughout the area, potable and non-potable (industrial) water, 
lighting, emergency lighting, ventilation and air conditioning. A building 
management system provides controls throughout the Campus. Cooling is 
provided with two redundant 50-ton (633~MJ) chillers supplying 
chilled water to three air handling units that provide ventilation to 
separate Campus spaces. Chilled water is also available for experiments to 
connect equipment directly. A dedicated 1500~kVA substation provides 
sufficient capacity for the experiment and facility needs, with margin for 
future expansion. Emergency power for lighting is provided with batteries 
in the lighting system to provide immediate light, while a standby diesel 
generator near the campus provides up to 24 hours of power to all safety 
systems in the campus. This includes water pumps in the nearby Yates shaft 
to prevent water from rising into the campus spaces.

While space is at a premium at the Davis Campus, two cutouts exist 
outside the clean space offering footprints of 33--50 m$^{2}$, with an 
average ceiling height of 3.7 m.  These spaces could be made available to 
groups with modest equipment needs.

\subsubsection{Rock Overburden}

When considering the muon flux at key laboratory spaces on the 4850L, six 
geological formations are important to consider: Grizzly, Flagrock, 
Northwestern, Ellison, Homestake and Poorman.  The Yates Member (Unit) is 
the lowest stratigraphic unit of the Poorman Formation and is important 
for developments on the 4850L.  In addition, tertiary Rhyolite dikes occur 
throughout the rock mass.

\begin{figure}
\includegraphics*[scale=0.35]{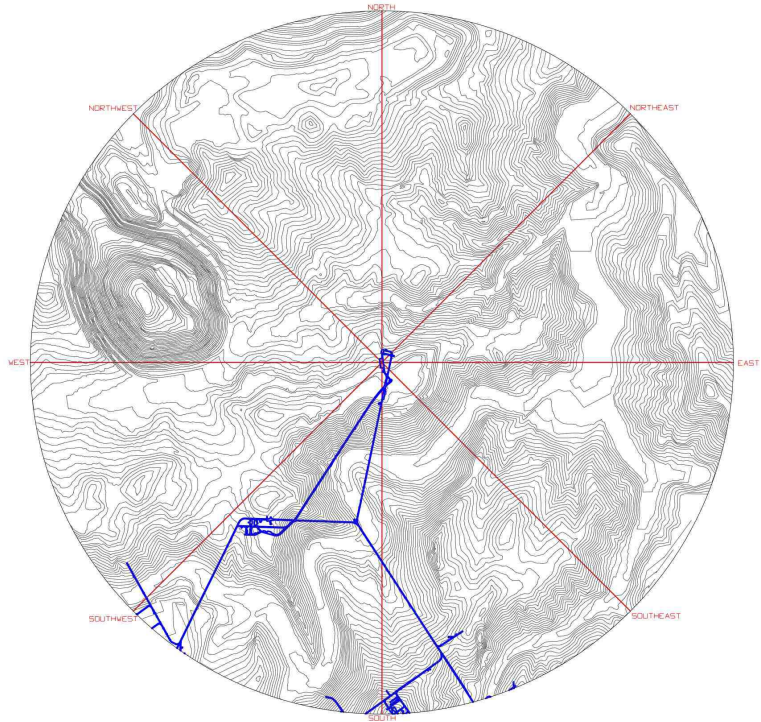}
\caption{\label{fig:Topo_4850L} Topological contours overlaid on top of an 
outline of the 4850L (shown in blue).  Various cardinal coordinates are 
indicated with red lines.  The Open Cut is a prominent feature on 
the surface toward the west.  Contour lines indicate 20-foot elevation 
changes.}
\end{figure}

The surface topology varies significantly throughout the laboratory 
property as shown in Figure~\ref{fig:Topo_4850L}.  Using average rock 
densities and preliminary geological model data in addition to recent 
topological surveys, an estimate of the rock overburden can be determined 
as shown in Table~\ref{tab:mwe}.  Overburden uncertainties will be 
quantitatively evaluated in an upcoming publication.  Geochemical 
composition data for various formations are also being compiled from 
sources such as the USGS~\cite{Caddey-1991} and others.

\begin{table}[htbp]
\caption{\label{tab:mwe}Rock overburden estimates for different 
underground laboratory locations at SURF, reflecting considerable 
variations in surface topology as well as geology.  Uncertainties on the raw 
overburden values are on the order of a few percent.}
\begin{tabular}{lcc}
\hline \hline
{\bf 4850L Laboratory}            & \multicolumn{2}{c}{\bf Rock Overburden} \\
{\bf Location}                    & {\bf (m)} & {\bf (m.w.e)} \\
\hline
Davis Campus -- LUX               & 1464   & 4200 \\ 
Davis Campus -- MJD               & 1478   & 4300 \\ 
Davis Campus -- Entrance Cutout   & 1561   & 4400 \\ 
Ross Campus -- MJD Electroforming & 1503   & 4300 \\ 
Ross Campus -- LBNE               & 1386   & 3900 \\ 
\hline \hline
\end{tabular}
\end{table}


\subsubsection{Cleanliness}

Cleanliness at the Davis Campus is maintained through a combination of 
transition zones and protocols as well as dedicated cleaning staff.  When 
supplies and equipment are brought from the dirty space into the clean 
space, items either need to be suitably enclosed or cleaned in the 
facility entrance Cart Wash.  Personnel typically enter though a series of 
rooms where outer coveralls and dirty-side gear is removed in favor of 
corresponding clean items.  Laundry services are also available in the 
clean space.

When the laboratory is occupied, particle counts in all common areas are 
below 10,000/ft$^{3}$.  The {\sc Majorana} collaboration 
follows additional gowning protocols to achieve ultra-low cleanliness 
levels in their laboratory spaces.  A summary of cleanliness states 
throughout various areas is presented in Table~\ref{tab:cleanliness}.  
The facility design includes two sets of showers, but so far they have not 
be necessary.


\begin{table}[htbp]
\caption{\label{tab:cleanliness}Median particle count levels in various 
laboratory locations during occupied periods.}
\begin{tabular}{lcc}
\hline \hline
{\bf Davis Campus Location}              & {\bf Particle Count} \\
                                         & {\bf (0.5~$\mu$m/ft$^{3}$)} \\
\hline
Transition, Laundry Area                 & 1290--8200 \\
Transition, Common Corridor              & 3160 \\
Davis Cavern, Lower                      & 3400 \\
Davis Cavern, Lower Counting Room        & 710 \\
Davis Cavern, Upper                      & 450  \\
Transition, {\sc Majorana} Detector Room & 100--400 \\
\hline \hline
\end{tabular}
\end{table}


\subsubsection{Radioactivity and Radon}

SURF strives to provide the lowest possible radioactivity environment for 
experiments hosted within the facility. This commitment had been 
integrated into the site preparation process from the early days of the 
facility design, and carried over to the realization of the 4850L Davis 
Campus laboratories.  These efforts include site and environmental 
characterization including rock radioactive measurements, use of 
low-radioactivity construction materials, and regular monitoring of 
environmental factors, including airborne radon.

The natural abundance of U/Th/K in Homestake rock is generally low, 
especially in certain geological formations such as the Yates Amphibolite, 
which have been measured to contain sub-ppm levels of U/Th.  Samples from 
other formations such as the Rhyolite intrusions can be 30--40x higher as 
illustrated in Table~\ref{tab:uthk}.  While not as low as the host 
Amphibolite rock, the lowest-activity construction materials for the Davis 
Campus were selected from a large number of samples 
~\cite{Assay-Whitepaper, Assay-Oroville, Assay-YDC}.


\begin{table}[htbp]
\caption{\label{tab:uthk} Partial U/Th/K assay results for Homestake rock 
samples as well as key construction materials used at the Davis 
Campus~\cite{Assay-Whitepaper, Assay-Oroville, Assay-YDC}.}
\begin{tabular}{lccc}
\hline \hline
              & {\bf Uranium} & {\bf Thorium}  & {\bf Potassium} \\ 
              & {\bf (ppm)}   & {\bf (ppm)}    & {\bf (\%)} \\
\hline
Amphibolite               & 0.22  & 0.33  & 0.96 \\
Rhyolite Dike             & 8.75  & 10.86 & 4.17 \\[0.2cm]
Shotcrete -- Low Activity & 1.52  & 2.17  & 0.55 \\
Shotcrete -- Standard     & 2.00  & 3.35  & 1.23 \\
Shotcrete -- Finish Coat  & 1.62  & 3.08  & 0.79 \\
Masonry Blocks            & 2.16  & 3.20  & 1.23 \\
\hline \hline
\end{tabular}
\end{table}


Long-term underground radon data have been collected since 2009 and at the 
Davis Campus since July 2012, shortly after outfitting was completed.  
Aside from the normal facility ventilation that brings fresh air via the 
Ross and Yates shafts, no extraordinary steps have been taken to mitigate 
underground radon levels.  Some research groups employ a type of purging 
system to reduce radon locally for their equipment.  As seen at other 
underground laboratories, a seasonal dependence is becoming apparent in 
the Davis Campus trends, with higher radon levels observed during the 
summer months May through September.  The total average radon 
concentration for all periods at the Davis Campus is approximately 340 
Bq/m$^{3}$ with a low baseline of 150 Bq/m$^{3}$ during the winter months.

Other efforts to characterize physics backgrounds (eg., muons, neutrons, 
gamma rays) in various underground areas were carried out by research 
groups over several years~\cite{Bkgd-Radon, Bkgd-Gamma, Bkgd-Muon}.  
Results from recent neutron flux measurements collected on the 4850L are 
expected in an upcoming publication.

\section{Current Science Program}

Science efforts that started in 2007 during re-entry into the facility 
have grown steadily over the years.  Building on the legacy of the Ray 
Davis chlorine solar-neutrino experiment~\cite{Davis} that began in 1965 
at the Homestake Mine, more than a dozen research groups have established 
underground research programs at SURF.  Measurements have been made and 
samples collected from elevations ranging from surface to the 5000-foot 
level, investigating topics in physics, geology, biology and engineering.  
Research programs that are considered to be currently active are listed in 
Table~\ref{tab:science_programs} as well their respective locations.  
Three main physics efforts are underway at SURF.

\subsection{Large Underground Xenon (LUX)}

The LUX experiment is conducting a direct search for weakly interacting 
massive particles (WIMPs) using 350~kg of purified xenon inside an 
ultra-pure titanium cryostat that is immersed in a water tank with 
272,550~liters (72,000~gallons) of purified water.  Following collisions 
in the liquid xenon volume, a strong electric field moves ionization 
electrons into the gas space at the top of the detector, and a total of 
122 PMTs collect scintillation light from interactions in both xenon 
volumes.

Members of the LUX collaboration have been active onsite at SURF since 
December 2009, when preparations began for detector assembly at the 
Surface Laboratory.  After completing detector assembly and operational 
testing on surface, LUX began their transition underground to the Davis 
Campus in May 2012.  With strong support from SURF personnel, the 
fully-assembled LUX detector was successfully transported underground from 
the Surface Laboratory in the Yates cage and into the Davis Cavern -- 
stringent limits on acceleration and tilt were not exceeded.  The LUX 
space at the Davis Campus includes a cleanroom and a control room, which 
is separate from the main lab space.  The SURF water purification system 
and a storage area for liquid nitrogen are located near the LUX equipment 
but outside the clean space.

The LUX collaboration recently published results from their first 
underground data run~\cite{LUX}.

\subsection{{\textsc Majorana Demonstrator} (MJD)}

The {\sc Majorana} collaboration is investigating neutrinoless double-beta 
decay using a detector called the {\sc Demonstrator} consisting 40 kg of 
germanium in two ultra-pure copper cryostats and a compact lead shield.  
{\sc Majorana} expects to use up to 30~kg of enriched $^{76}$Ge.  The main 
technical goal of the {\sc Demonstrator} is to confirm that the ambitious 
purity requirements for a tonne-scale detector are achievable~\cite{MJD}.  
The group also hopes to test a controversial claim for the detection of 
neutrinoless double-beta decay.

In order to satisfy the very low background requirements, the {\sc 
Majorana} collaboration has set up a laboratory at the 4850L Ross Campus 
to produce the world's purest electroformed copper, which is then machined 
at a dedicated machine shop at the Davis Campus.

The {\sc Majorana} collaboration began work onsite starting in November 
2010 with deliveries of equipment (including the first germanium 
detectors) and then with preparations for the electroforming cleanroom.  
Production of electroformed copper began in July 2011 and about eight 
months later the group started to move equipment into the Davis Campus.

\subsection{Center for Ultra-low Background Experiments in the Dakotas (CUBED)}

CUBED is a South Dakota Governor's Research Center that is administered 
through the University of South Dakota and involves physics researchers 
and others from the majority of universities in the state.  In addition to 
increasing the South Dakota academic involvement in SURF research, the 
center is maintaining a focus on areas of interest congruent with planned 
SURF experiments, such as crystal growth, low-background counting and 
isotope separation.

A low-background counting laboratory has been set up in a dedicated room 
in the Lower Davis Cavern.  The system allows direct-gamma counting using 
a 1.2-kg high-purity germanium crystal, resulting in 0.01--0.1~ppb 
sensitivities to U/Th.  Initial operation has begun, and production 
counting is expected to start in early 2014 once the outer shielding is 
complete.

Another CUBED project that proposes to investigate isotopic separation and 
ultra-purification techniques is being considered for the Surface 
Laboratory.  Funding is being sought to develop a laboratory to support 
underground crystal growth.


\newpage
\begin{table}[htbp]
\caption{\label{tab:science_programs}Current scientific research programs 
at the Sanford Laboratory. Locations in bold indicate current 
installations or the subject of current focus.}
\begin{tabular}{llll}
\hline\hline
\multicolumn{1}{c}{\bf Experiment} & \multicolumn{1}{c}{\bf Description} & 
\multicolumn{1}{c}{\bf Locations}  & \multicolumn{1}{c}{\bf References} \\
\hline
\multicolumn{4}{l}{\bf Physics} \\
\hline
LUX & Dark matter using Xe & Surface, {\bf 4850L} & \cite{LUX} \\[0.1cm]
{\sc Majorana Demonstrator} & Neutrinoless double-beta decay using Ge & 
{\bf Surface}, {\bf 4850L} & \cite{MJD} \\[0.1cm]
CUBED & Low-bkgd counter; also bkgds: & 
{\bf Surface}, 800L, & \cite{CUBED}, \cite{Bkgd-Radon}, \cite{Bkgd-Gamma}, 
\cite{Bkgd-Muon} \\ 
(also Bkgd Characterization) & muon, neutron, gamma, radon & 2000L, 
4550L, {\bf 4850L} &\\[0.1cm]
DIANA & Neutron bkgds & 4100L, 4850L & \footnote{Publication in progress.}\\[0.1cm]
LBNE & Cleanliness tests & Surface, 4850L & \cite{LBNE-Tiedt} \\[0.1cm]
Deep Underground Gravity & Seismic characterization for gravity-wave 
& Surface, 300L, & \cite {DUGL-2010_1}, \cite{DUGL-2010_2}\\
Laboratory (DUGL) & research & 800L, 2000L, 4100L &\\
\hline
\multicolumn{4}{l}{\bf Geology} \\
\hline
Geoscience Optical Extensometers & Optical fiber apps, tiltmeters & {\bf 2000L}, 
{\bf 4100L}, & \cite{Geo-GEOXTM-2013}, \cite{Geo-GEOXTM-2012_1}, 
\cite{Geo-GEOXTM-2012_2}, \cite{Geo-GEOXTM-2011}, \cite{Geo-GEOXTM-2010} \\
and Tiltmeters (GEOX$^{\rm TM}$) && {\bf 4850L} &\\[0.1cm]
USGS Hydrogravity & Local gravity for water tables, densities & 
{\bf Surface}, {\bf 4100L}, {\bf 4850L} & \cite{Hydrogravity} \\[0.1cm]
Petrology, Ore Deposits, & Core archive and logs, geologic mapping & 
Surface, 800L &  \cite{PODS-Armstrong}, \cite{PODS-Hamer} \\
Structure (PODS) &&& \\[0.1cm]
Transparent Earth & Seismic monitoring & 2000L, {\bf 4100L} & 
\cite{Geo-TransparentEarth-2013},\cite{Geo-TransparentEarth-2011},\cite{Geo-TransparentEarth-2009}\\ 
\hline
\multicolumn{4}{l}{\bf Biology} \\
\hline
Biodiversity (BHSU/SDSMT) & Microbiology & Surface, 300L, 2000L, & \cite{Bio-Waddell-2010} \\
&& {\bf 4100L}, 4550L, 4850L &\\[0.1cm]
Biofuels (SDSMT) & Biofuels & 4550L, {\bf 4850L}, 5000L & 
\cite{Bio-Bhalla-2013_1}, \cite{Bio-Rastogi-2013_1}, 
\cite{Bio-Bhalla-2013_2}, \cite{Bio-Rastogi-2013_2}, 
\cite{Bio-Rastogi-2010_1}, \\
&&& \cite{Bio-Rastogi-2010_2}, \cite{Bio-Rastogi-2009_1}, \cite{Bio-Rastogi-2009_2} \\[0.1cm]
Bioprocessing R\&D (SDSMT) & Biofuels & {\bf 4850L} & \\[0.1cm] 
Syngas/Biofuels (SDSMT) & Biofuels & {\bf 4850L} & 
\cite{Bio-Shende-2012_1}, 
\cite{Bio-Shende-2012_2}, \cite{Bio-Shende-2012_3} \\[0.1cm]
Life Underground -- & Water in drill holes, geomicrobiology & Surface, 
{\bf 800L}, & \\
NASA Astrobiology Institute && 1700L, 2000L, {\bf 4850L} &\\
\hline
\multicolumn{4}{l}{\bf Engineering~\footnote{No active engineering 
research currently, but there are interested groups.  Previous groups 
investigated electromagnetic signal propagation in rock tunnels and 
autonomous submersibles.}} \\
\hline
\hline \hline
\end{tabular}
\end{table}


\newpage

\section{Future Science}

\begin{figure*}
\includegraphics*[scale=0.15]{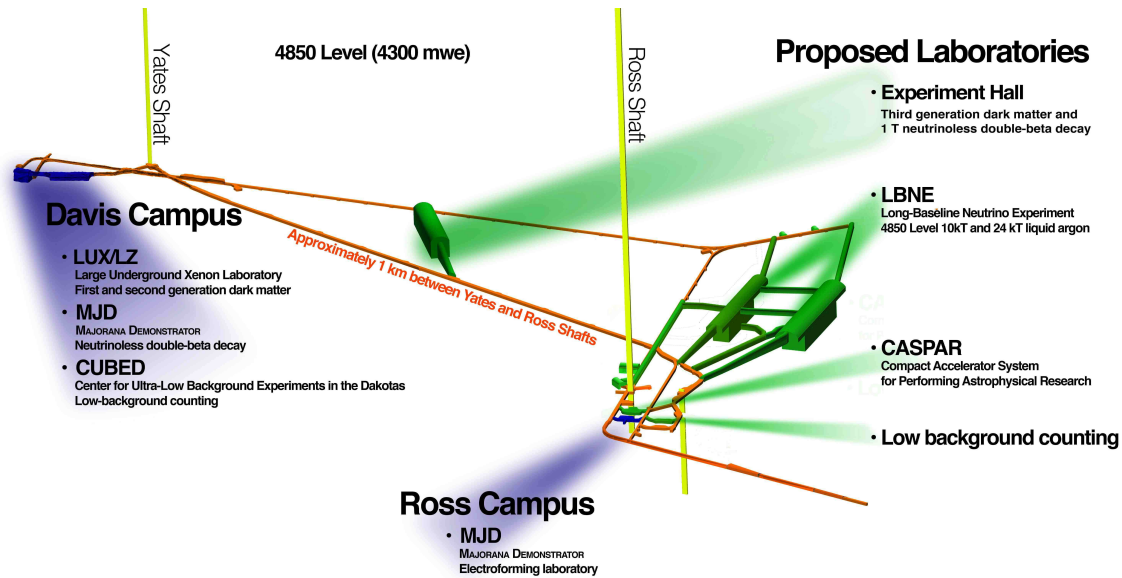}    
\caption{\label{fig:4850LWithCASPAR} The 4850 level of SURF highlighting 
the existing and proposed experiments.}
\end{figure*}

The are many promising options to expand the number of projects both in 
the near-term and into the next decade.  The locations of existing and 
future experiments is shown in Figure~\ref{fig:4850LWithCASPAR}.

\subsection{Low-Background Counting Facility}

Several proposals have been submitted proposing to develop 
state-of-the-art low background counting detectors and facilities at the 
4850L at SURF making use of existing infrastructure and/or excavations.  
One possible location for a low-background counting facility is in the SE 
shop space at the 4850L Ross Campus.

\subsection{CASPAR/DIANA}

As part of the NSF support for the development of underground physics 
experiments, a proposal has been developed to create an underground 
accelerator facility for low-energy nuclear astrophysics experiments. The
DIANA project (Dual Ion Accelerators for Nuclear Astrophysics) would 
consist of two high-current accelerators offering a low-energy beam 
(30--400~kV) and a high-energy beam (350--3000~kV)~\cite{DIANA}.

Planning has begun for the first phase of the project known as the Compact 
Accelerator System for Performing Astrophysical Research (CASPAR).  It 
will consist of a 1-MV Van de Graff accelerator that will be re-located 
from the University of Notre Dame.  Installation in the NE shop of the 
4850L Ross Campus could begin by the end of 2014 with commissioning in 
2015.

\subsection{Next Generation Dark Matter}

The DOE and NSF have plans to develop Generation-2 dark matter 
experiments, with masses up to $\sim$10~tonnes. The LUX-ZEPLIN (LZ) 
collaboration has submitted a proposal to develop a G-2 experiment in SURF 
using the existing space and infrastructure in the Davis Campus.  
Construction could begin as early as 2016 followed by commissioning in 
2017.  SURF is open to working with all interested collaborations to 
develop proposals for G-2 experiments.

The DOE has discussed developing a ``roadmap'' for dark matter searches 
including Generation-2 and Generation-3 dark matter experiments beginning 
$\sim$FY17. Initial efforts to develop the Generation-3 collaboration are 
advancing.

\subsection{Neutrinoless Double-Beta Decay}

The {\sc Majorana} collaboration is investigating whether it would be 
possible to perform the next-generation neutrinoless double-beta decay 
experiment on the 4850L at SURF.  If the combination of rock overburden 
and veto was determined to be sufficient, a new laboratory may be needed.

\subsection{Long Baseline Neutrino Experiment (LBNE)}

The LBNE experiment offers a compelling science program that will enhance 
the our understanding of the neutrino in several fundamental 
ways~\cite{LBNE}.  A large liquid argon detector (or detectors) would 
observe neutrinos generated at Fermilab using an upgraded accelerator beam 
(up to 2.3~MW).  The detector mass could be staged (eg., 10~kt, 17~kt, 
34~kt, etc).

Designs have been developed for an initial detector deployment on the 
surface, but the new planning baseline focuses on the installation 
underground on the 4850L.  The reduced backgrounds due to cosmic ray 
muons~\cite{Bkgd-Muon-LBNE} significantly enhances the physics reach of 
the LBNE detector.  Construction may start late 2017 with transition to 
operations by 2023.

\section{Summary}

SURF is a deep underground research facility, dedicated to scientific 
uses.  Research activities are supported at a number of facilities, both 
on the surface and underground.  Two campuses on the 4850L accommodate a 
number of leading efforts, and in particular the 4850L Davis Campus has 
been successfully operating for 18 months.  The LUX and {\sc Majorana} 
experiments are well established and there will soon be capabilities for 
low-background counting.  Many expansion possibilities are on the horizon 
and a number of key experiments in the U.S.\ research program are 
developing plans for installation at SURF.


\bibliographystyle{aipproc}   

\bibliography{Heise}

\end{document}